\documentclass[utf8]{FrontiersinHarvard}
\usepackage[T1]{fontenc}
\usepackage[english]{babel}
\makeatletter
\expandafter\let\csname l@en\endcsname\l@english
\expandafter\let\csname captionsen\endcsname\captionsenglish
\expandafter\let\csname dateen\endcsname\dateenglish
\makeatother

\usepackage{url,hyperref,lineno,microtype}
\usepackage[onehalfspacing]{setspace}

\DeclareUnicodeCharacter{03B1}{$\alpha$}
\usepackage{tipa}
\DeclareUnicodeCharacter{025B}{\textepsilon}

\newcommand{\revision}[1]{{{#1}}}

\def\firstAuthorLast{McKevitt {et~al.}}

\def\Authors{%
James McKevitt\,$^{1,2,*}$, Ignacio Ugarte-Urra\,$^{3}$ and Peter R. Young\,$^{4,5}$%
}
\def\Address{%
$^{1}$Mullard Space Science Laboratory, University College London, Holmbury St Mary, Dorking, Surrey, United Kingdom \\
$^{2}$Institute of Astrophysics, University of Vienna, Vienna, Austria \\
$^{3}$Space Science Division, Naval Research Laboratory, Washington, DC, USA \\
$^{4}$Solar Physics Laboratory, Heliophysics Science Division, NASA Goddard Space Flight Center, Greenbelt, MD, USA \\
$^{5}$Department of Mathematics, Physics and Electrical Engineering, Northumbria University, Newcastle upon Tyne, United Kingdom%
}
\def\corrAuthor{James McKevitt}

\def\corrEmail{james.mckevitt.21@ucl.ac.uk}

\begin{document}
\onecolumn
\firstpage{1}

\title[Hinode/EIS Full-Disk \revision{Spectroscopy} Across a Solar Cycle]{Full-Disk \revision{Spectroscopy} of the Solar Corona Across a Solar Cycle with Hinode/EIS}

\author[\firstAuthorLast ]{\Authors}
\address{}
\correspondance{}
\extraAuth{}

\maketitle

\begin{abstract}

\revision{
The structure and dynamics of the solar corona evolve with the Sun's magnetic cycle, yet how this variability manifests in the disk-integrated, Sun-as-a-star observables used in stellar activity studies remains poorly constrained. We compile 18 full-disk spectroscopic mosaic scans made by Hinode/EIS spanning 2013-2024, covering solar cycle 24 and the rise of solar cycle 25, and probe coronal plasma variability through the integrated and spatially resolved intensity, Doppler velocity, and non-thermal velocity of log~T$\sim$6.2 plasma in active regions and the quiet Sun. Disk-integrated coronal intensity is strongly correlated with the solar cycle, consistent with stellar observations. No clear solar-cycle variation is found in the distributions of coronal Doppler or non-thermal velocity in either active regions or the quiet Sun, though their total intensities do track the cycle. Active region intensity per unit solid angle shows a moderate correlation with solar cycle. Taken together, these results support the hypothesis that Sun-as-a-star coronal intensity variability across the solar cycle is driven primarily by the changing fraction of the disk occupied by active regions, rather than by changes in the log~T$\sim$6.2 plasma properties of those regions. The well-established correlation between upflowing plasma and elevated non-thermal line broadening in active regions persists throughout the cycle, implying that the underlying kinematic properties of active region plasma are insensitive to the global magnetic field configuration, a result with direct implications for the interpretation of coronal activity cycles on solar-like stars.
}

\tiny
 \section{Keywords:} solar corona, solar cycle, EUV spectroscopy, full-disk spectroscopy, active regions%
\end{abstract}

\section{Introduction}

The solar corona is structured by the Sun's magnetic field, which evolves over the 11-year solar cycle. During solar minimum, the global magnetic field is dominated by a low-order, largely axisymmetric dipolar component, giving rise to long-lived polar coronal holes and an enhanced fraction of open magnetic flux. In contrast, solar maximum is characterised by a far more complex and multipolar topology driven by the frequent emergence of active regions at low and mid-latitudes, which disrupts the large-scale dipole and redistributes open flux across the solar surface \citep[e.g.,][]{mackay_suns_2012, babcock_topology_1961}. These shifts in magnetic topology govern the global behaviour of coronal plasma, dictating the distribution of open versus closed field lines and, consequently, the global balance of plasma confinement and outflows.
Extreme ultraviolet (EUV) spectroscopy provides direct diagnostics of these dynamics. Doppler shifts trace bulk motions along the line of sight, while excess line broadening (often expressed as a non-thermal velocity) can reflect unresolved wave activity, turbulence, or multi-component flows \citep[e.g.,][]{doschek_flows_2008}. In spatially resolved solar observations, coronal holes are commonly associated with blue-shifted emission indicative of outflows \citep[e.g.,][]{hassler_solar_1999}, whereas active regions show a mix of red-shifts in cooling loops and blueshifts at footpoints and peripheries \citep[e.g.,][]{tian_upflows_2021}. Non-thermal broadening is also frequently enhanced in regions exhibiting upflows, although the physical drivers are not understood, in part due to instrumental limitations \citep[e.g.,][]{doschek_dynamics_2012}.

\revision{

Stellar spectra are spatially unresolved and combine emission from multiple magnetic structures \citep[e.g.,][]{kowalski_stellar_2024}, so stellar atmospheric dynamics cannot be studied with solar-like spatial detail. Sun-as-a-star observations, where emission from the solar disk is spatially integrated (as in stellar spectra) but where the spatially-resolved contributors to the emission are known, can be useful in understanding how disk-integrated spectra represent the atmospheres of stars \citep{toriumi_bridging_2026}. \citet{toriumi_sun-as--star_2020}, for example, showed that integrated multi-wavelength emission retains information on the presence and evolution of active regions, plages, and coronal loops. One complication is that stellar activity studies usually use chromospheric diagnostics such as Ca II H\&K, \citep[e.g.,][]{costes_long-term_2021} whereas resolved solar coronal studies often use optically thin EUV lines \citep[e.g.,][]{mckevitt_coronal_2026}. Nevertheless, on long timescales chromospheric and coronal diagnostics broadly trace the same underlying cycle in stars, and solar magnetic flux correlates with atmospheric emission from the chromosphere to the corona \citep{toriumi_universal_2022}.

In these solar-stellar contexts solar atmospheric irradiance is typically measured using full-disk spatially-resolved broadband imaging, given the availability of such data from the Atmospheric Imaging Assembly onboard the Solar Dynamics Observatory (SDO/AIA; \citealp{pesnell_solar_2012,lemen_atmospheric_2012}). 
The Extreme ultraviolet Variability Experiment on SDO (EVE; \citealp{woods_extreme_2012}) provides full-disk Sun-as-a-star spectra, but cannot isolate contributions from specific solar regions. 
No observatory 
currently provides 
sustained, high-cadence, spatially-resolved full-disk spectral diagnostics of the solar atmosphere \citep{ugarte-urra_case_2023}. Small-field-of-view spectrographs such as the EUV Imaging Spectrometer onboard Hinode (Hinode/EIS; \citealp{kosugi_hinode_2007,culhane_euv_2007}) provide unmatched plasma diagnostics, but typically cover less than 10\% of the full solar disk so can miss large-scale coupling and distant destabilisation \citep{schrijver_pathways_2013,schrijver_statistical_2015}
. 
Therefore full-disk spectroscopic studies require mosaics assembled from sequences of individual observations.

Coronal activity cycles are known 
on stars \citep[e.g.,][]{hempelmann_coronal_2006,robrade_coronal_2012}. The `Sun as an X-ray star' series \citep[e.g.,][]{orlando_sun_2000} used solar differential emission measure (DEM) analysis from the Soft X-ray Telescope on Yohkoh \citep{tsuneta_soft_1991,ogawara_solar-mission_1991} to infer 
the fraction of stellar disks occupied by active regions (filling factor)%
. \citet{orlando_sun_2001} found that active region filling factor is the main cause of 
solar-cycle emission-measure-variability%
. Also, 
\citet{morgan_global_2017} used broadband SDO/AIA imaging to show that solar-cycle-related variability in full-disk emission is driven by the presence of active regions, but that active regions themselves display no clear solar-cycle-related variability in average temperatures, emission measures, or magnetic field strength. Such work supports the hypothesis that stellar coronal activity cycles are caused by active region filling factor. However, approaching this with spectroscopic data, rather than broadband imaging, would prove useful in challenging the hypothesis further.}

\revision{To this end, we compile and analyse a set of 18 spatially resolved full-disk spectroscopic mosaics taken between 2013 and 2024 by Hinode/EIS, processed using the \texttt{EISMaps} pipeline \citep{mckevitt_coronal_2026}, and perform an analysis of the variability of log~$T\left[\mathrm{K}\right]\sim$6.2 plasma. We also release full-disk maps of intensity, Doppler velocity, and non-thermal velocity for log~$T\left[\mathrm{K}\right]\sim$4.7--7.1 plasma, which were processed but whose analysis fell outside the scope of this study.
}

\section{Observations and Data Reduction}

\revision{Hinode/EIS performs different full-disk mosaic observations using both it's narrow slits \citep[e.g.,][]{brooks_full-sun_2015}, and it's slots\footnote{Available under the Hinode Operations Plan (HOP) 0130: \url{https://www.isas.jaxa.jp/home/solar/hinode_op/hop.php?hop=0130}} \citep[e.g.,][]{warren_absolute_2014}. In this study we use 18 full-disk spectroscopic narrow-slit mosaic scans, taken between 2013 and 2024, for which Level-1 data are available\footnote{\url{https://eis.nrl.navy.mil/}}.} To enable the full-disk scan to be completed within a typical 48-hour Hinode observing window, the widest 2~arcsec slit is used and the scan step size is set to 4~arcsec (sparse rastering). The plate scale of 1~arcsec results in spectrally-resolved measurements of plasma in pixels of 4~arcsec (X) $\times$ 1~arcsec (Y).

\subsection{Spectroscopy}

Due to the time taken to make the full disk scan, the exposure time for each observation is limited, resulting in only a few emission lines possessing resolved spectra with a signal-to-noise (SNR) ratio sufficient to derive reasonable non-thermal velocity values from fitted emission lines. In this study we fit and analyse the strong Fe~XII~195.119~\AA{}. We find the other coronal lines, Fe~XI~188.216~\AA{} and Fe~XIII~202.044~\AA{}, to be sufficiently strong to derive reliable non-thermal velocity values across the disk, while all lines captured in the \revision{Hinode/EIS observations} show intensity and Doppler velocity values \revision{with signal to noise ratios sufficient to analyse solar-cycle trends.}

\revision{We assemble a Sun-as-a-star spectrum for each disk by summing the spatially-resolved spectra after correcting for the various instrument effects, as discussed in detail in the Appendix. We also make spatially-resolved full-disk maps of intensity, velocity, and non-thermal velocity. To do this, }we fit Gaussian profiles to the \revision{spectra}
using the \texttt{MPFIT} algorithm \citep{markwardt_non-linear_2009} implemented in the EIS Python Analysis Code \citep[EISPAC;][]{weberg_eispac_2023}. \revision{In the case of Fe~XII~195.119, }we use two-component Gaussian fits to account for the blended Fe~XII~195.179~\AA{} line. \revision{Other lines are fitted with multi-component Gaussian templates of EISPAC where blends are present.} The Doppler velocity \(v\) was calculated from the fitted line centroid \(\lambda\), rest wavelength \(\lambda_0\), and speed of light \(c\) using \(v=c\frac{\lambda-\lambda_0}{\lambda_0}\). We normalise measured values for their position on disk using $v_{\text{norm}} = v/{\cos{\theta}}$, where $\theta$ is the angle between the line of sight and the normal to the solar surface. The excess broadening (non-thermal velocity; $v_{nt}$) was calculated using

\begin{equation}
    {\text{FWHM}_{o}}^2={\text{FWHM}_{i}}^2+4\ln{2}\left(\frac{\lambda}{c}\right)^2\left({v_t}^2+{v_{nt}}^2\right),
\end{equation}

\noindent{}where $\text{FWHM}_{o}$ and $\text{FWHM}_{i}$ refer to the observed and instrumental full width at half maximum values respectively, and where $v_t$ is the thermal velocity \revision{calculated using the peak-ionisation-fraction temperature of the ion}. We take the instrumental width from \texttt{eis\_slit\_width} in the EIS software tree in Solarsoft. Additionally, we correct intensity measurements using the radiometric calibration of \cite{del_zanna_hinode_2025}, and individual rasters making up the full disk map are assembled using the method of \cite{mckevitt_coronal_2026}\footnote{Here we use version 0.1.2: \url{https://doi.org/10.5281/zenodo.17641183}. The source code and ongoing development are available at: \url{https://github.com/jamesmckevitt/eismaps}}. We make the processed FITS files for the disks available\footnote{\url{https://doi.org/10.5522/04/31304956}}.

\subsection{Solar features}

We use Space-weather HMI Active Region Patches \citep[SHARP;][]{bobra_helioseismic_2014} from the Helioseismic and Magnetic Imager onboard SDO \citep[SDO/HMI;][]{scherrer_helioseismic_2012}, to locate active regions. We use multi-wavelength intensity imaging from SDO/AIA and the neural network method of \cite{jarolim_multi-channel_2021} to locate coronal holes. We then define quiet Sun, for the purpose of this study, as pixels on the disk not captured by active region or coronal hole masking. \revision{For our spatially-resolved measurements, }we additionally exclude pixels \revision{with heliocentric angle $\theta>$85\textdegree{}, where $\theta$ is the angle between the observer's line of sight and the local solar surface normal. This corresponds to pixels lying within 5\textdegree{} of the solar limb on the visible disk, where the large Doppler cosine correction amplifies artifacts.}
We do not consider the fitted parameters in coronal holes in our analysis given their low intensity further reduces signal-to-noise and their susceptibility to interference from other effects \citep[see e.g. scatter light effects discussed by][]{young_scattered_2022}.

\section{Results}

\subsection{Variability of plasma distributions}

\begin{figure}
    \centering
    \includegraphics[width=\linewidth]{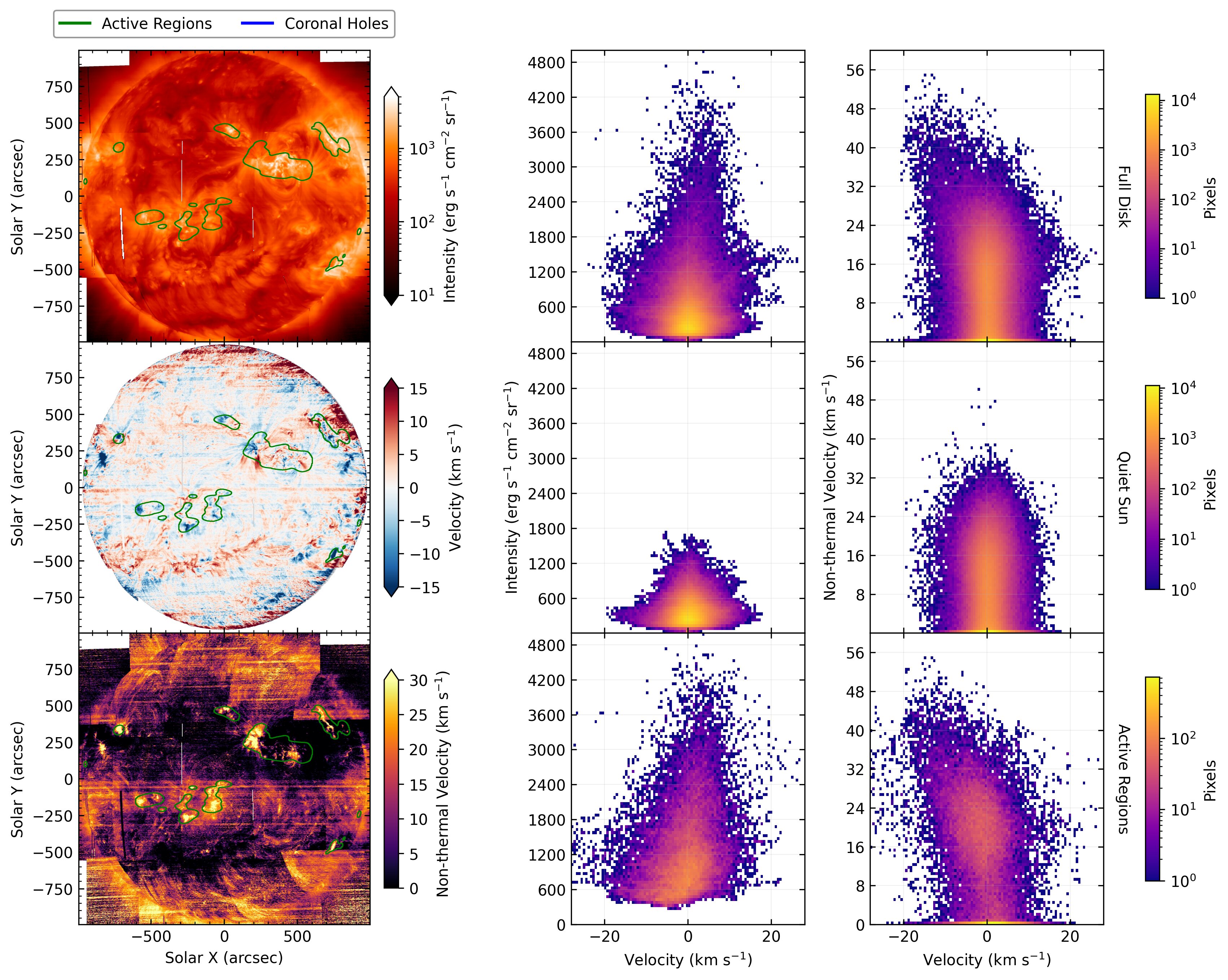}
    \caption{\revision{Left: The first spectroscopic full-disk mosaic taken by Hinode/EIS where the panels, from top to bottom, show intensity, velocity, and non-thermal velocity. Right: Histograms of intensity, velocity and non-thermal velocity, for the full disk, the quiet Sun, and active regions. Active regions and coronal holes are outlined in green and blue respectively.}}
    \label{fig:fig_firstdisk}
\end{figure}

\begin{figure}
    \centering
    \includegraphics[width=\linewidth]{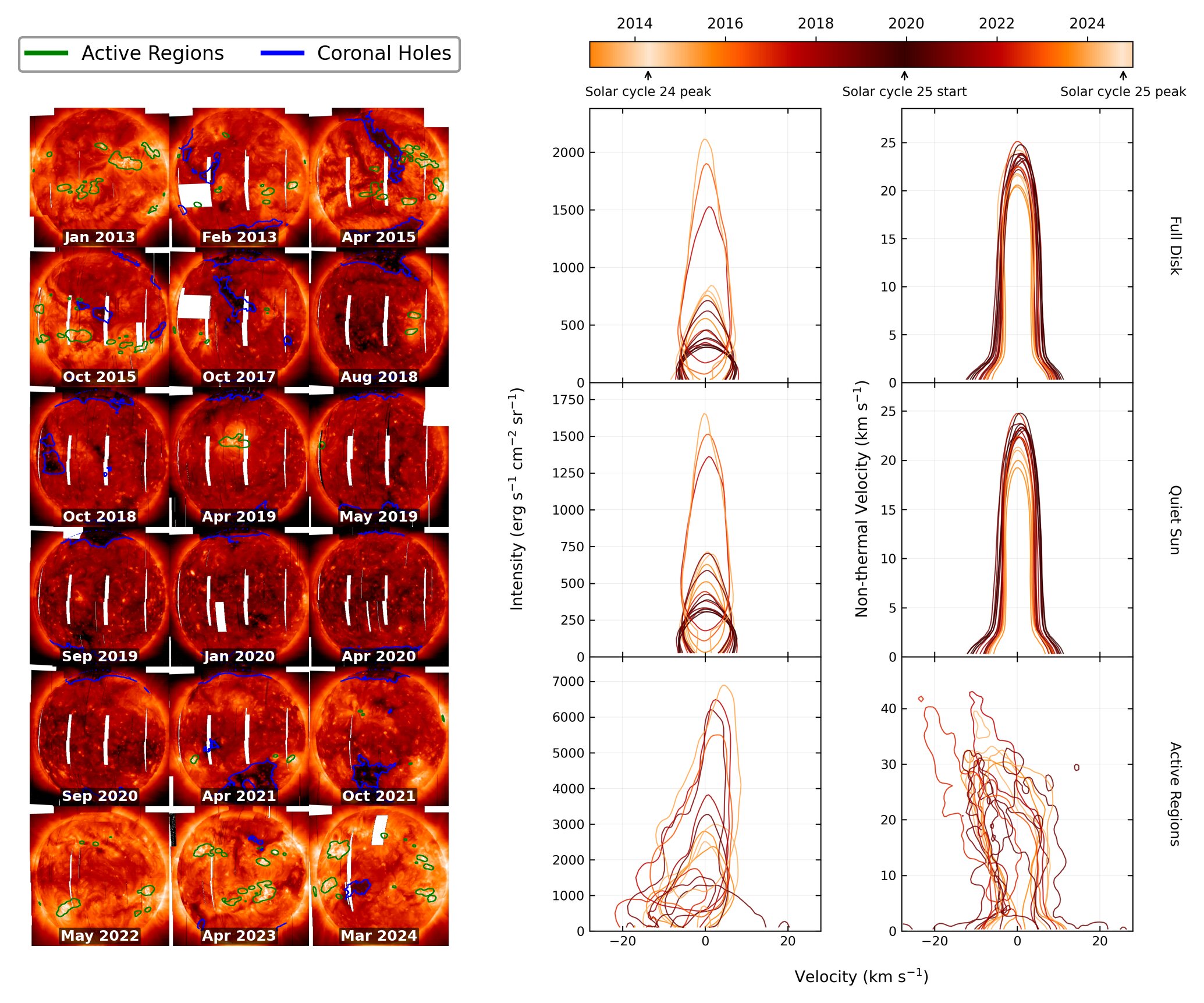}
    \caption{Left: Full-disk spectroscopic mosaics taken by Hinode/EIS used in this study, with active region and coronal hole masks outlined \revision{in green and blue respectively}. Right: Distributions of plasma intensity, velocity, and non-thermal velocity at different points in the solar cycle, identified using contours at 75\% of the histogram density for each disk.}
    \label{fig:all_disk_distributions}
\end{figure}

\revision{
Figure~\ref{fig:fig_firstdisk} shows measurements of log~$T\left[\mathrm{K}\right]\sim$6.2 coronal plasma from the first disk in the series with maps of coronal intensity, velocity, and non-thermal velocity on the left, and 2D histograms of velocity versus intensity and non-thermal velocity for the full disk, quiet Sun, and active regions on the right. Our histograms use bins of 50~erg/s/cm$^2$/sr in intensity, 0.56~km/s in velocity, and 0.60~km/s in non-thermal velocity. Figure~\ref{fig:all_disk_distributions} shows the corresponding distributions for all 18 disks, represented by 75\% histogram-density contours coloured by solar-cycle phase. We take the peak of solar cycle 24 as April 2014, the start of solar cycle 25 as December 2019, and the peak of solar cycle 25 as October 2024 \citep{clette_silso_2015}. These contours capture the core plasma distributions well. The only excluded feature in this first disk is a sparse tail of high full-disk non-thermal velocities in upflowing plasma, which may be associated with active regions. Given the low pixel counts in this tail and possible velocity systematics from instrumental spectral drift (see Appendix), we do not consider it further in this study.
}

\subsubsection{Doppler Velocity and Intensity}

The distribution of the full disk plasma shows that the spread of velocities remains consistent throughout the solar cycle. The primary variation is a shift in peak intensity; during solar maximum, the distribution peaks at higher intensities but the velocity centroid remains stable at $\sim$0~km/s. The quiet sun shows a similar distribution to the full disk, though with a reduced peak intensity. 
Active regions display a strong dependence on intensity. At the highest intensities, Doppler velocities are tightly clustered around $\sim$0~km/s and even slightly red-shifted. However, as intensity decreases, the distribution skews significantly towards upflowing plasma. This relationship appears largely consistent across the solar cycle.

\subsubsection{Doppler and Non-thermal Velocities}

For the full disk, the distribution is symmetric around zero Doppler velocity, with no clearly distinguishable variations with the solar cycle. We find the same to be \revision{largely} true for the quiet Sun distribution, \revision{but note some disks near to solar maximum show slightly lower peak non-thermal velocities}. 
In active regions, stronger non-thermal velocities are correlated with upflowing plasma. There is no clear variation in this relationship with respect to the solar cycle.


\subsection{Disk-integrated variability}

\begin{figure}
    \centering
    \includegraphics[width=\linewidth]{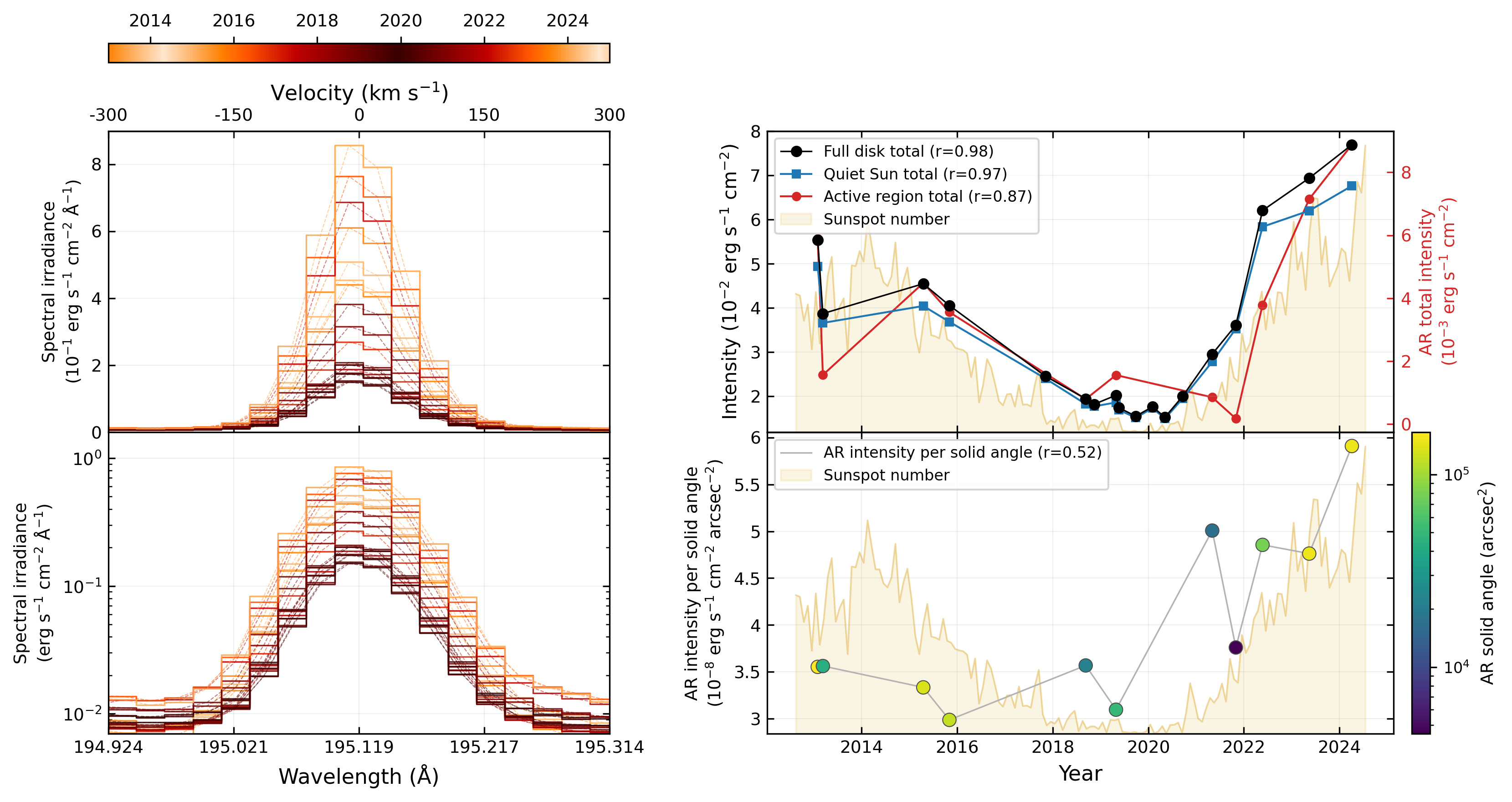}
    \caption{\revision{Left: Sun-as-a-star spectra around Fe~XII~195.119 at different points in the solar cycle, assembled from Hinode/EIS full-disk mosaics, with linear and logarithmic intensity axes (top and bottom panels respectively). Right: The total intensity of the full disk, quiet Sun, and active region plasma for each disk in the series (top). The sunspot number is shown to illustrate the solar cycle, and the correlation coefficients between the intensities and the sunspot number are shown in the legend. The active region intensity normalised by solid angle for each of the disks, with the solar cycle (bottom). Points are coloured according to the total active region solid angle. The sunspot number and correlation coefficient are also shown. The active region intensity for the October 2017 and May 2019 disks are excluded due to their small solid angle and incidence with bad data (see Figure~\ref{fig:all_disk_distributions}).}}
    \label{fig:fig_sun_as_star_combined}
\end{figure}


\revision{To consider our results in a Sun-as-a-star context, we show the integrated spectra around Fe~XII~195.119 at different cycle phases in the left panels of Figure~\ref{fig:fig_sun_as_star_combined}. The 
lines display similar profiles, but their total intensity varies with the solar cycle, with stronger emission near solar maximum.
}
\revision{
The top-right panel shows the Sun-as-a-star full-disk, quiet-Sun, and active-region line intensities with the solar cycle together with sunspot number, where all three are strongly positively correlated with the cycle (r=0.98, 0.97, and 0.87). 
The bottom-right panel shows active-region intensity normalised by active-region solid angle, coloured by total active-region solid angle. 
Missing pixels in some disks were corrected by assigning them the mean intensity of the same feature type elsewhere on the disk, to ensure all disks had the same number of contributing pixels. We find a moderate correlation between active-region intensity per solid angle and the solar cycle (r=0.52), with higher values during the rise of cycle 25.
}

\section{Discussion}

\revision{
This study finds that disk-integrated coronal EUV intensity tracks the solar cycle closely (r=0.98; Figure~\ref{fig:fig_sun_as_star_combined}), and that the distributions of Doppler and non-thermal velocity within the magnetic structures responsible for that emission show no clear modulation across the same interval. This strong cycle dependence at the integrated level, and near-invariance at the per-structure level, supports the hypothesis that active region filling factor is the primary driver of coronal intensity variability, rather than any change in the spectroscopic characteristics of the regions themselves. We note, however, that spectroscopic characteristics such as composition and thermal structure were not covered in this study. \citet{orlando_sun_2001} reached a similar conclusion using analysis of soft X-ray DEM variability between the maximum and the minimum of solar cycle 22. \citet{morgan_global_2017} subsequently supported this result using SDO/AIA broadband imaging across most of cycle 24 and some of cycle 25, demonstrating that the average temperature, emission measure, and photospheric magnetic field strength of individual active regions showed no systematic cycle-phase dependence. The spectroscopic analysis we present is sensitive to plasma motions and wave activity in a way that broadband imaging and DEM analysis is not, and so further supports the filling factor hypothesis.
}

The persistence of the well-known \revision{anti-correlation} between upflows and non-thermal broadening in active regions throughout the cycle implies that the mechanism(s) responsible \revision{(e.g. impulsive footpoint heating, unresolved Alfv\'enic wave activity, or reconnection-driven outflow)} operate whenever the magnetic topology of an active region is present, regardless of the global solar magnetic field. \revision{We caution, however, that our sample size of active region plasma at solar minimum is limited.} 
\revision{The moderate correlation between active region intensity normalised by solid angle with the solar cycle (r=0.52; Figure~\ref{fig:fig_sun_as_star_combined}) is therefore curious. A pure filling-factor model would predict no residual correlation once solid angle is accounted for. The fact that one remains at moderate significance suggests that either the typical active region near solar maximum carries higher intrinsic intensity, perhaps reflecting higher photospheric flux and more complex magnetic geometry, or that diffuse emission from the outskirts of active regions, not fully captured by the SHARP-based masking, contaminates the solid angle normalisation. The former would be consistent with the known dependence of coronal luminosity on total unsigned magnetic flux, which depends not only on the area occupied by magnetic structures but also on their field strength \citep{pevtsov_relationship_2003}. In that sense, normalising only by active region solid angle would not necessarily remove all solar-cycle-dependent variability. 
Distinguishing between enhanced unsigned magnetic flux per unit active region area near solar maximum and contamination from diffuse emission at active region outskirts would require more directly comparing the spectroscopic dataset against HMI magnetic flux, a step that falls outside the scope of the present analysis.}

\revision{There is some small variation seen with solar cycle in the peak non-thermal velocity values in the quiet Sun plasma distributions, where solar maximum is associated with slightly lower non-thermal velocities. This could be due to stronger emission reducing artificial enhancers of non-thermal velocity such as signal-to-noise ratio and stray light effects. However, intensity versus non-thermal velocity histograms would be required to properly consider this, something outside the scope of this study. The next-generation high-throughput SOLAR-C/EUVST spectrometer will enable such measurements to a much higher precision \citep{mckevitt_pre-flare_2026}.} 
\revision{The quiet-Sun distributions appear to show variability resembling that of the full disk. We find this logical given the majority of the disk in all cases is quiet Sun plasma. However, we note that the active region masking technique we apply may also influence the quiet Sun distributions, where diffuse or extended emission from nearby active regions is considered as quiet Sun plasma.}

The original motivation for assembling this dataset was to connect spatially resolved coronal spectroscopy to Sun-as-a-star observables, and in particular to ask whether any solar-cycle variability in disk-integrated line profiles can be explained by changes seen within specific magnetic structures. \revision{Modelling of X-ray activity cycles in solar analogues \citep[e.g.,][]{hempelmann_coronal_2006,robrade_coronal_2012,orlando_fifteen_2017,coffaro_x-ray_2020} has consistently considered varying active region filling factors as the mechanism linking chromospheric and coronal cycles, with per-region plasma treated as more fixed. That has rested largely on the assumption that coronal loop physics is determined locally. Our results suggest, albeit with limited data points, that the rise phase of the solar cycle and the related magnetic complexity may complicate this relationship.
}



\revision{

\section*{Appendix: Limitations of Sun-as-a-Star measurements from Hinode/EIS}

Creating a Sun-as-a-star spectrum requires summing the spatially-resolved spectra measured by Hinode/EIS into one single spectrum. This spectrum is then quasi-equivalent to one taken which integrates emission from the whole disk at once, as is done in observations of stars. However, assembling such a spectra requires a different approach to the one used to assemble our spatially-resolved full disks. 
A single coherent three-dimensional datacube of those two spatial and one common spectral axis, must be assembled, which can then be summed into a Sun-as-a-star spectra. Due to various instrument effects seen on Hinode/EIS, pixels do not natively share a common spectral axis in either the scan or slit direction, and are affected by other effects to different degrees. This complicates the assembly of such Sun-as-a-star spectra, which we approach in this appendix.

\subsection{Emission line drift}

Due to the thermal deformation of the instrument, which changes during the orbit of the spacecraft and within the time taken for one single raster scan, optical components move relative to one another and cause spectral lines to drift across the detector with time. This causes artificial Doppler shifts of up to 70~km/s \citep{kamio_modeling_2010}. The method of \citet{kamio_modeling_2010} produces an absolute wavelength calibration correction for this effect. However, in recent years the performance of this correction has degraded. Therefore, EISPAC implements an additional correction whereby the median Doppler velocity measured along the slit is assumed to be 0~km/s at each slit position. The median measured velocity per slit position is then used to apply an offset to all measured velocities. One further complication is a tilt in the slits relative to the detectors \citep{young_eis_2010}, which needs to be accounted for. The standard corrections for these effects broadly function well. However, these corrections mean that the spectral bins used to resolve the emission lines change position relative to the line's rest wavelength for each image pixel, so that no common spectral axis is present
. Therefore, in our Sun-as-a-Star assembly, we resample all pixels to a common wavelength grid
. We do this by first fitting the spatially resolved spectra and determining the median 0~km/s offset in the spectral axis for each slit position, then applying this offset to the raw spectra in wavelength and resampling all spectra to a common wavelength axis using the integrated flux-conserving spectral resampler of \citet{carnall_spectres_2017}, as implemented by \citet{the_astropy_collaboration_astropy_2022}.

\subsection{Degraded detector pixels}

Some pixels on Hinode/EIS have been degraded during the lifetime of the instrument due to the high-energy particle environment in space. These are detected and corrected using interpolation \citep{young_interpolating_2010}. We apply the same correction procedure in our Sun-as-a-Star pipeline.



\subsection{Instrumental broadening}

Recovering the broadening of the emission line is more complicated. The primary limiting factor is the instrument's point spread function (PSF). The broadening of spectral lines induced by the optical system varies with position along the slit \citep{young_instrumental_2011}. Typical processing of Hinode/EIS data, such as we show in Figure~\ref{fig:fig_firstdisk} in our spatially-resolved measurements, first fits the broadened emission lines to calculate an observed line width and then subtracts the estimated instrumental width in quadrature. This results in a width parameter which is the quadrature sum of the line's thermal and non-thermal broadening components \citep[e.g.][]{mckevitt_link_2024}. In the case of Hinode/EIS, instrumental broadening forms the majority of the line width, and varies by $\sim$30\% along the slit for the 2~arcsec slit used for full-disk observations (see Figure~1 of \citealt{young_instrumental_2011}). To assemble the required coherent 3D datacube we are required to remove the bias introduced into some image pixels by this effect, normalising all pixels to their true thermal and non-thermal widths. This means deconvolving the variable slit width from the individual measured spectra.

\begin{figure}
    \centering
    \includegraphics[width=0.5\linewidth]{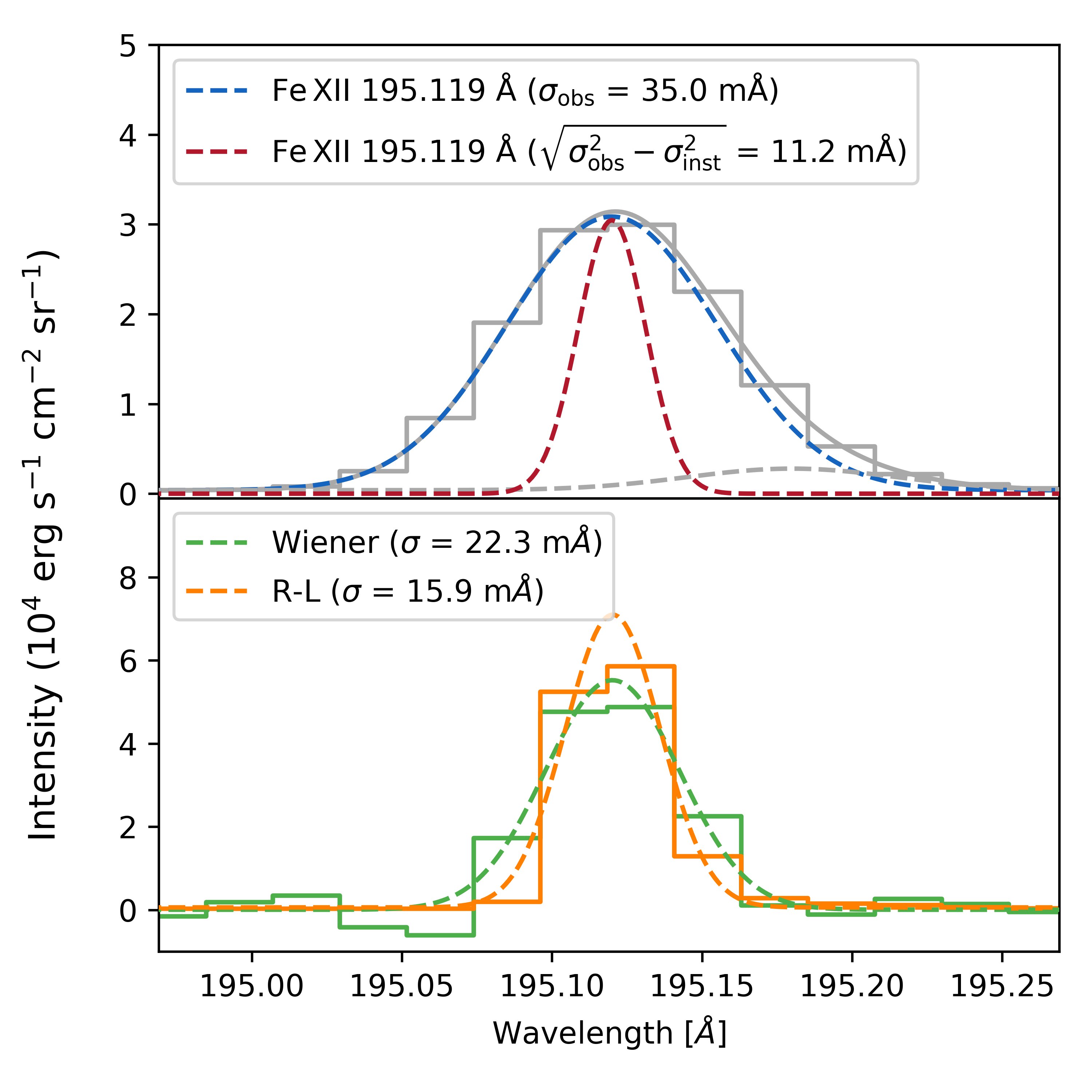}
    \caption{95th percentile intensity pixel of the first raster in the first Hinode/EIS full disk. Top panel shows the observed spectra and best fit two-component Gaussian in solid grey, the component corresponding to the Fe~XII~195.119~\AA{} in dashed blue and that same component with the instrumental width removed in dashed red. We also show the fitted Fe~XII~195.179~\AA{} component in dashed grey. Bottom panel shows attempts to deconvolve the slit width directly from the spectra in solid lines, and a best-fit single-component Gaussian in dashed lines.}
    \label{fig:deconv_spectra}
\end{figure}

We approached this problem by first performing the standard quadrature-subtraction method, and then comparing the performance of different deconvolution methods to produce the same width. To do this, using the first raster scan of the first full disk, we selected the pixel with intensity nearest the 95th percentile and extracted the Fe~XII~195.119~\AA{} spectra. We fit this spectra using the standard EISPAC two-component Gaussian to capture the blended Fe~XII~195.179~\AA{} emission line. We then subtracted the instrumental width from the primary 195.119~\AA{} component to find the `true' width of the emission line (top panel Figure~\ref{fig:deconv_spectra}). Separately, we applied the Wiener and Richardson-Lucy deconvolution algorithms to the raw spectra \citep{wiener_extrapolation_1949,richardson_bayesian-based_1972,lucy_iterative_1974}, and fitted the resulting emission lines to measure their widths (bottom panel Figure~\ref{fig:deconv_spectra}). The results shown in the bottom panel of Figure~\ref{fig:deconv_spectra} were for Wiener $\varepsilon$=0.05 and Richardson-Lucy $n$=100. We varied $\varepsilon$ between 0.01 and 0.1 and $n$ from 10 to 1000, but found no improvements in performance. As is seen in the figure, the deconvolution cannot successfully remove the instrumental width.
To better understand this we explored the performance of our deconvolution approaches on synthetic spectra in a parameter space of varying instrumental width and spectral sampling rates, and found that where instrumental broadening is larger than the intrinsic width of the emission line, deconvolution with these approaches struggles. We conclude that, given on Hinode/EIS the majority of the total line width is generated by the instrument PSF, recovery of the spectra using deconvolution is challenging. We note, however, that recent results of \citet{brosius_nonthermal_2025} suggest EIS instrumental widths may be narrower than those presented by \citet{young_instrumental_2011}.

Due to these limitations we did not remove instrumental broadening from the emission lines before their assembly. As each disk is observed with the same pattern of observations, covering each portion of the disks almost identically, each disk's spectra is affected the same by this instrumental broadening. Therefore, the spectra are normalised against this instrumental affect and comparable to each other. It should also be noted that instrumental broadening does not affect the total intensity of the spectra, such as we consider in Figure~\ref{fig:fig_sun_as_star_combined}, and so does not affect our analysis. 

Other approaches which could be considered to normalise spectra might be to widen the lines where instrumental broadening contributes less, or to sum the individually fitted Gaussians with instrumental width removed to create a Sun-as-a-star Gaussian. In the latter case, care would need to be taken to conserve integrated intensity (see Figure~\ref{fig:deconv_spectra}).


}

\section*{Conflict of Interest Statement}
The authors declare that the research was conducted in the absence of any commercial or financial relationships that could be construed as a potential conflict of interest.

\section*{Author Contributions}

JM: Conceptualization, Data curation, Formal analysis, Funding acquisition, Methodology, Software, Visualization, Writing - original draft; IUU: Conceptualization, Formal analysis, Methodology, Resources, Writing - review \& editing; PY: Formal analysis, Methodology, Resources, Writing - review \& editing.

\section*{Funding}
J.M. was supported by STFC PhD Studentship number ST/X508858/1. I.U.U. acknowledges support from the NASA Hinode programme. P.R.Y. acknowledges funding from the GSFC Internal Scientist Funding Model competitive work package programme, and the Hinode project.

\section*{Acknowledgments}
\revision{We thank the reviewers for their thoughtful and thorough comments, suggestions, and insight, which greatly improved this manuscript. We thank Prof. Manuel Güdel for his insightful comments on the solar-stellar connection. GitHub Copilot was used to assist with code review.} Data analysis was performed using Austrian Scientific Computing infrastructure (https://asc.ac.at/). Hinode is a Japanese mission developed and launched by ISAS/JAXA, with NAOJ as domestic partner and NASA and STFC (UK) as international partners. It is operated by these agencies in co-operation with ESA and NSC (Norway).

\section*{Data Availability Statement}
The datasets generated and analysed in this study can be found in the UCL Research Data Repository \url{https://doi.org/10.5522/04/31304956}.

\bibliographystyle{Frontiers-Harvard}
\bibliography{references}

\end{document}